\documentclass[12pt,preprint]{aastex}
\usepackage{amssymb}
\usepackage{amsmath}
\usepackage{epstopdf}
\usepackage{natbib}
\usepackage[colorlinks, citecolor=blue]{hyperref}
\usepackage{hyperref}
\usepackage[all]{hypcap}

\begin{document}

\title{Neutrino Emission in Jet Propagation Process}
\author{D. Xiao\altaffilmark{1,2} and Z. G. Dai\altaffilmark{1,2}}
\affil{\altaffilmark{1}School of Astronomy and Space Science, Nanjing University, Nanjing 210093, China; dzg@nju.edu.cn}
\affil{\altaffilmark{2}Key Laboratory of Modern Astronomy and Astrophysics (Nanjing University), Ministry of Education, China}

\begin{abstract}

Relativistic jets are universal in long-duration gamma-ray burst (GRB)
models. Before breaking out, they must propagate in the progenitor
envelope along with a forward shock and a reverse shock forming at
the jet head. Both electrons and protons will be accelerated by the
shocks. High energy neutrinos could be produced by these protons
interacting with stellar materials and electron-radiating photons.
The jet will probably be collimated, which may have a strong effect
on the final neutrino flux. Under the assumption of a power-law
stellar-envelope density profile $\rho \propto r^{-\alpha}$ with an
index $\alpha$, we calculate this neutrino emission flux by these
shocks for low-luminosity GRBs (LL-GRBs) and ultra-long GRBs (UL-GRBs)
in different collimation regimes, using the jet propagation framework
developed by \citet{bro11}. We find that LL-GRBs and UL-GRBs are
capable for detectable high energy neutrinos up to $\sim {\rm PeV}$,
and obtain the final neutrino spectrum. Besides, we conclude
that larger $\alpha$ corresponds to greater neutrino flux at high
energy end ($\sim {\rm PeV}$) and higher maximum neutrino energy as well.
However, such differences are so small that it is not promising
for us to distinguish from observations, given the energy resolution
we have now.

\end{abstract}


\section{Introduction}

The collapsar model of gamma-ray bursts suggests that during
the core collapse of a progenitor massive star to a neutron star or
a black hole, a relativistic jet punctures the stellar envelope and
transports energy to electrons and protons through shock
acceleration \citep{zha04, pir05, mes06, woo93, mac99}. Accelerated
electrons dominate radiation via synchrotron or inverse Compton
mechanism while accelerated protons produce neutrinos by
proton-proton collision and photopion process \citep{wax97, wax99,
rac98, alv00, bah01, gue03, mura06, bec08}. Within this scenario, we
can expect high energy neutrinos originating from different stages
in a GRB event. First, \citet{wax00} and \citet{dai01} proposed
$\epsilon_\nu>10^3\,{\rm TeV}$ neutrinos from an external reverse
shock. Similarly, neutrinos from an external forward shock have
been discussed \citep{li02, der03, raz13}. Second, the prompt
photon emission from GRBs can be correlated to the production of
neutrinos, since protons are also believed to be accelerated in
relativistic internal shocks \citep{vie95, wax97, wax99,
mes00, gue04}. \citet{mur06} obtained a diffuse neutrino background
spectrum from GRBs for specific parameter sets in the internal shock
model. Alternatively, the neutrino emission might arise in a
dissipative jet photosphere \citep{ree05, mur08, wan09, gao12}.
Third, neutrinos could also be produced while the
jet is still propagating in the envelope. Generally, these neutrinos
appear as a precursor burst. \citet{mes01} predicted a neutrino
precursor of $ \epsilon_\nu \geq 5\,{\rm TeV}$, which is produced by
internal shocks at radius $r_{\rm IS} \approx 10^{10}-10^{11}\,{\rm
cm}$. \citet{enb09} presented a new analysis of the neutrino flux of
these internal shocks for two types of source environments: the slow-jet
supernova (SJS) model and the GRB model. Still further,
\citet{mur13} studied high energy neutrino production in collimated
jets inside progenitors of gamma-ray bursts and supernovae,
considering both collimation and internal shocks. \citet{pru03}
discussed the neutrinos produced by inelastic neutron-nucleon
collisions of a relativistic jet propagating through a stellar
envelope. \citet{raz03} dealt with the high energy neutrino
signature in a supernova remnant shell ejected prior to a gamma-ray
burst and then \citet{and05} extended their model and significantly
improved the detection prospects. \citet{hor08} proposed that a
reverse shock at the jet head probably will accelerate protons when
crossing the jet material, and they calculated in detail the
cumulative neutrino event number that may be observed by a ${\rm
km}^2$ scale detector like IceCube.

The jet propagation dynamics has been studied. Assuming a constant
jet velocity, \citet{beg89} discussed the propagation of a galactic
jet in the intergalactic medium. \citet{mes01} analyzed the
propagation in the envelope of a red supergiant star, ignoring the
surrounding cocoon. \citet{mat03} studied the jet-cocoon structure
and jet head velocity that is constrained by ram pressure
balance, but they did not consider the collimation by cocoon
pressure. \citet{laz05} took into account the collimation effect and
meanwhile they assumed that the jet expands adiabatically.
\citet{bro11} considered a collimated shock that forms at the base
of the jet and dissipates parts of the jet's energy to
counterbalance the cocoon's pressure. They figured out the geometry
of a collimated shock, and thus further obtained the requirement for
jet collimation. Moreover, it is worth mentioning that various
numerical simulations of jet propagation in the envelope already
have been performed \citep{zha03, mor07, miz09, miz13}.

In our paper we try to calculate the flux of high energy neutrinos
from the shocks formed at the jet head for low-luminosity GRBs (LL-GRBs)
and ultra-long GRBs (UL-GRBs), when the jet is still propagating
inside the envelope. Most importantly, we take into account the jet
collimation which could largely affect the final neutrino flux
but was ignored in the previous studies. For simplicity, we only
account for jet propagation in the helium core because collimation
mainly happens there. We assume a power law envelope density profile
$\rho(r) =A r^{-\alpha}$, where $A=(3-\alpha)M_{\rm He}/(4\pi r_{\rm
He}^{3-\alpha})$ and $2\leq \alpha < 3$ with $M_{\rm He}$ and
$r_{\rm He}$ being the mass and radius of the helium envelope. We
employ the analytical solutions described in \citet{bro11} to
further calculate the neutrino flux in different collimation
regimes and at last we discuss its dependence on the index $\alpha$,
providing an alternative way to probe the GRB progenitor
through neutrino precursor signal in the future.

This paper is organized as follows. In section 2 we discuss the jet
propagation dynamics. In section 3 we perform detailed calculations
of neutrino flux in each regime and a comparison between
different regimes. Dependence on $\alpha$ is considered in section
4. We make a comparison with previous works in Section 5 and finally
we finish with discussions and conclusions in section 6.

\section{Jet Propagation Dynamics}

According to \cite{mat03}, the head velocity is constrained by ram
pressure balance,
$$\rho_jh_j\Gamma_j^2\Gamma_h^2(\beta_j-\beta_h)^2+P_j=\rho_ah_a\Gamma_h^2\beta_h^2+P_a, \eqno (1)$$
where $\rho, P, \beta, \Gamma$ and $h\equiv1+4P/\rho c^2$ are the
density, pressure, velocity, Lorentz factor and dimensionless
specific enthalpy, and the subscripts $j$ and $a$ refer to jet and
ambient material. Then the jet head velocity is
$$\beta_h=\frac{\beta_j}{1+\tilde{L}^{-1/2}},\eqno (2)$$
with
$$\tilde{L}\equiv\frac{\rho_jh_j\Gamma_j^2}{\rho_a}\simeq\frac{L_j}{\Sigma_j\rho_ac^3}, \eqno (3)$$
where $L_j$ is the jet luminosity and $\Sigma_j=\pi r_j^2$ is the jet's cross section.
\par
The parameter $\tilde{L}$ is crucial for determining collimation
regimes \citep[see][Table 1]{bro11}. While $\tilde{L} \lesssim
\theta_0^{-4/3}$ the jet is strongly collimated by the cocoon
pressure, where $\theta_0$ is the jet opening angle. \cite{bro11}
would further divide it into two situations: $\tilde{L} \ll 1 $ and
$1\ll \tilde{L} \lesssim \theta_0^{-4/3}$. The
uncollimated regime corresponds to $\tilde{L} > \theta_0^{-4/3}$, accordingly.

The internal energy and particle number density of the shocked and
unshocked regions are correlated by \citep{bla76, sar95}
$$\frac{e_f}{n_fm_pc^2}=\Gamma_h-1,     \quad     \frac{n_f}{n_a}=4\Gamma_h+3,$$
$$\frac{e_r}{n_rm_pc^2}=\bar{\Gamma}_h-1,  \quad   \frac{n_r}{n_j}=4\bar{\Gamma}_h+3, \eqno
(4)$$ where the subscripts $f$ and $r$ represent regions that have
been crossed by the forward shock and reverse shock. $\Gamma_h$ is
the Lorentz factor of the head and $\bar{\Gamma}_h$ is the Lorentz
factor of the unshocked jet measured in the jet head frame,
$$\bar{\Gamma}_h=\Gamma_j\Gamma_h(1-\beta_j\beta_h). \eqno (5)$$
Moreover, the density of the unshocked jet materials $\rho_j$ is
determined by
$$ L_j=\Gamma_j^2\pi r_j^2\rho_jc^3. \eqno (6)$$
Based on the above equations, we can carry on our calculation.

\section{Theoretical Calculation of Neutrino Flux}

In this section, we take $\alpha=2$ as our premise, and discuss
the dependence on $\alpha$ later in section 4.
\par
The most promising acceleration process in a GRB is Fermi
acceleration mechanism. But now there are noteworthy arguments
that once the the jet bulk kinetic energy is dissipated
at the collimation shock, the collimated jet would become
radiation dominated and then the reverse shock occurring
at the interface of the jet head and collimated jet would
also be radiation mediated. At such shocks, photons produced
in the downstream diffuse into the upstream and interact with
electrons or pairs. There would not be a strong shock jump
any more and Fermi acceleration no longer works \citep{lev08, kat10, mur13}.
However, this case is changed for LL-GRBs \citep{sod06, tom07, lia07, mur13}
and UL-GRBs \citep{lev13, gen13, mur13}. Because of low power or large radii,
the Thomson optical depth is low even inside a star \citep{mur13}
so that efficient Fermi acceleration would be expected. We consider
LL-GRBs and UL-GRBs separately and they fall into different
collimation regimes, which will be talked about later.
\par
Due to the first order Fermi acceleration, the accelerated proton
spectrum is \citep{ach01, kes05, hor08}:
$$\frac{{\rm d}n_p}{{\rm d} \epsilon_p} \propto \epsilon_p^{-p}, \eqno (7)$$
where $\epsilon_p$ and $n_p$ are the proton energy
and number density. We optimistically take $p=2$ in our
calculation. The minimum proton energy $\epsilon_{p,\min}\sim
\Gamma_jm_pc^2$ and maximum energy $\epsilon_{p,\max}$ is determined
by the balance between proton acceleration and cooling process.

\subsection{$\tilde{L} \ll 1 $}
In this regime, the jet head moves forward with a non-relativistic
velocity, $\Gamma_h\simeq 1$.  The unshocked jet's Lorentz factor in
the head frame is $\bar{\Gamma}_h=\Gamma_j$. We can easily see that
the reverse shock is strong while the forward shock is so weak that
we need not consider. The internal energy of the shocked jet is
$e_r\simeq (\Gamma_j-1)(4\Gamma_j+3)\rho_jc^2$. In addition, we
deduce some crucial terms below analytically \citep[see][Appendix
B]{bro11}.
$$\tilde{L}=\left(\frac{16}{\pi}\right)^{2/3}L_j^{2/3}A^{-2/3}\theta_0^{-8/3}c^{-2},$$
$$\theta_j=2^{-4/3}\pi^{-1/6}L_j^{1/6}A^{-1/6}\theta_0^{8/15}c^{-1/2},$$
$$\rho_j=2^{8/3}\pi^{-2/3}\Gamma_j^{-2}L_j^{2/3}A^{1/3}r^{-2}\theta_0^{-16/15}c^{-2}, \eqno (8)$$
where $\theta_0$ is the initial jet opening angle and $\theta_j$ is the
opening angle after collimation. Fortunately, $\tilde{L}, \theta_j$
do not vary with radius when and only when $\alpha=2$.
\par
The internal energy
$$e_r=2^{8/3}\pi^{-2/3}\frac{(4\Gamma_j+3)(\Gamma_j-1)}{\Gamma_j^2}L_j^{2/3}A^{1/3}r^{-2}\theta_0^{-16/15}. \eqno
(9)$$
We assume the energy equipartition factor
$\varepsilon_e=\varepsilon_B=0.1$ and thus we can get the comoving
magnetic field by
$$\frac{B^2}{8\pi}=\varepsilon_Be_r,$$
$$B=2^{17/6}\pi^{1/6}\sqrt{\varepsilon_B\frac{(4\Gamma_j+3)(\Gamma_j-1)}{\Gamma_j^2}}L_j^{1/3}A^{1/6}r^{-1}\theta_0^{-8/15}. \eqno
(10)$$ Because of the large opacity of the envelope, the radiation of relativistic electrons will be thermalized, with a typical black body temperature
$$\frac{8\pi^5(kT)^4}{15(hc)^3}=\varepsilon_ee_r,$$
$$kT=2^{2/3}\pi^{-1/6}[\frac{15\varepsilon_e(hc)^3}{8\pi^5}\frac{(4\Gamma_j+3)(\Gamma_j-1)}
{\Gamma_j^2}]^{1/4}L_j^{1/6}A^{1/12}r^{-1/2}\theta_0^{-4/15}.\eqno (11)$$
The average number density of thermal photons is
$$n_\gamma=19.232\pi\times\frac{1}{(hc)^3}\times(kT)^3. \eqno (12)$$

\par
Furthermore, in order to get the numerical values, we adopt the typical
values of LL-GRBs, $L_{\rm iso}=10^{46}{\rm erg\,s}^{-1}$, $\theta_0=0.02$, which suggests $L_j=10^{42}{\rm erg\,s}^{-1}$. So the Lorentz factor of collimated jet is $\Gamma_j\sim 1/\theta_0\sim50$\citep{miz13}. According to \citet{heg00} and \citet{woo06}, we assume a typical progenitor with a helium core of mass $\sim 2M_\odot$ and radius $r_{\rm He}=4\times10^{11}\,$cm
so that the ambient envelope density can be expressed as $
\rho_a(r)=7.96\times 10^{20}r^{-2}\;{\rm g\,cm}^{-3}$. From equation (8) we get $\tilde{L}\simeq 0.013 \;,\;\rho_j=4.95\times 10^{-11}\;{\rm g\,cm}^{-3}$ so that this LL-GRB meets the requirement $\tilde{L} \ll 1 $. Moreover, Fermi acceleration is efficient because the Thomson optical depth is $\tau_T=n\sigma_T l= \frac{\rho_j}{m_p}\sigma_T\frac{r_{\rm He}}{\Gamma_j} \simeq0.158(\frac{L_j}{10^{42}{\rm erg\,s}^{-1}})^{\frac{2}{3}}(\frac{\theta_0}{0.02})^{\frac{29}{15}}(\frac{M_{\rm He}}{2M_\odot})^{\frac{1}{3}}(\frac{r_{\rm He}}{4\times10^{11}\rm cm})^{-\frac{4}{3}} \sim 0.1C^{-1}\bar{\Gamma}_h < 1$, where $C=1+2\ln\bar{\Gamma}_h^2$ is the possible effect by pair production \citep{mur13, bud10, nak12}. Now we can get $B=3.33\times10^7$G,
$kT=0.76\,{\rm keV}$, $n_\gamma=1.36\times10^{22}{\rm cm}^{-3}$.
Remember that this is measured in the rest frame of the jet head.
\par
A high energy proton loses its energy through radiative and hadronic
process.  The radiative cooling includes synchrotron and inverse
Compton scattering, with typical cooling timescales \citep{hor08}:
$$t_{\rm sync}=\frac{6\pi m_p^4c^3}{\sigma_Tm_e^2B^2\epsilon_p},$$
$$t_{\rm IC,Th}=\frac{3m_p^4c^3}{4\sigma_Tm_e^2\epsilon_\gamma n_\gamma \epsilon_p},$$
$$t_{\rm IC,KN}=\frac{3\epsilon_\gamma\epsilon_p}{4\sigma_Tm_e^2c^5n_\gamma}, \eqno (13)$$
where the subscripts ${\rm Th, KN}$ represent the Thomson limit and
Klein-Nishina limit of inverse Compton scattering, $\sigma_T$ is the
Thomson cross section and average photon energy
$\bar{\epsilon}_\gamma=2.7kT$. If the proton energy $\epsilon_p$ is in
units of GeV, the numerical values of the timescales given above are
$$t_{\rm sync}=t_{\rm IC,Th}=4.07\times10^{3}\frac{1}{\epsilon_p}\;{\rm s},\;\;\;t_{\rm IC,KN}=2.16\times10^{-8}\epsilon_p\;{\rm s}.$$
Hadronic cooling mechanisms mainly contain proton-proton collision,
the Bethe-Heitler interaction and photopion production
in the following ways respectively,
$$p+p\longrightarrow\pi^{\pm}, K^{\pm}\longrightarrow\mu^{\pm}+\nu_\mu(\bar{\nu}_\mu),$$
$$p+\gamma\longrightarrow p+e^{\pm},$$
$$p+\gamma\longrightarrow p+\pi^{0}\quad {\rm or} \quad p+\gamma\longrightarrow n+\pi^+.$$
We can expect muon neutrinos to be produced in pp and photopion
process, via charged pion or kaon decay. Here we do not consider the
secondary electron neutrino production.
\par
The cooling timescale of pp collision is
$$t_{pp}=\frac{\epsilon_p}{c\sigma_{pp}n_p\Delta{\epsilon_p}}.\eqno (14)$$
We estimate the proton number density in the shocked jet as
$n_p=(4\Gamma_j+3)\rho_j/m_p\sim 6.02\times 10^{15}{\rm cm}^{-3}$.
Assuming in each collision a fraction 20\% of the proton energy is lost and
$\sigma_{pp}=5\times 10^{-26}{\rm cm}^2$ \citep{eid04}, we can get
$t_{pp}=0.55\,{\rm s}$.
\par
At relative higher energy, the protons start to cool through the
Bethe-Heiter interaction,  for which the energy loss every times is
$\Delta{\epsilon_p}=2m_ec^2\gamma_{\rm c.m.}$, where $\gamma_{\rm
c.m.}$ is the Lorentz factor of the center of inertia in the
comoving frame and can be expressed as $\gamma_{\rm
c.m.}=(\epsilon_p+\epsilon_\gamma)/(m_p^2c^4+2\epsilon_p\epsilon_\gamma)^{1/2}$.
The BH cross section is given by $\sigma_{\rm BH}=(28/9)\alpha
r_e^2\ln[(2\epsilon_p\epsilon_\gamma)/(m_pm_ec^4)-106/9]$, so the BH
cooling time is \citep{raz04, hor08}
$$t_{\rm BH}=\frac{\epsilon_p}{2n_\gamma c\sigma_{\rm BH}m_ec^2\gamma_{\rm c.m.}}. \eqno (15)$$
\par
The photopion production dominates the cooling even at higher
energy.  We adopt the photopion cross section described in
\citet{ste68} and \citet{asa05} as a broken power law:
$\sigma(\chi)=5\times10^{-28}(\chi/590)^{3.2}{\rm cm}^2$ for
$290<\chi<590$ and
$\sigma(\chi)=5\times10^{-28}(\chi/590)^{-0.7}{\rm cm}^2$ for
$590<\chi<9800$, where $\chi m_ec^2$ is the photon energy in the
proton rest frame. Thus the cooling timescale is
$$t_{p\gamma}=\frac{\epsilon_p}{c\sigma_{p\gamma}n_\gamma\Delta{\epsilon_p}}, \eqno (16)$$
where the conventional inelasticity
$K=\Delta{\epsilon_p}/\epsilon_p=[1-(m_p^2-m_\pi^2)/s]/2$ and $s$ is
the invariant mass of the system.
\par
The timescale of the first order Fermi acceleration is $t_{\rm
acc}=\theta_F\epsilon_p/(eBc)$. If the diffusion
coefficient is assumed to be proportional to the Bohm diffusion coefficient and
$\theta_F$ is taken to be the common value
$\theta_F=10$ \citep{raz04, ando05, hor08}, then $t_{\rm acc}\sim
3.32\times10^{-11}\epsilon_p\,{\rm s}$.
\par
\begin{figure}[htbp]
\centering
\includegraphics[width=0.7\textwidth]{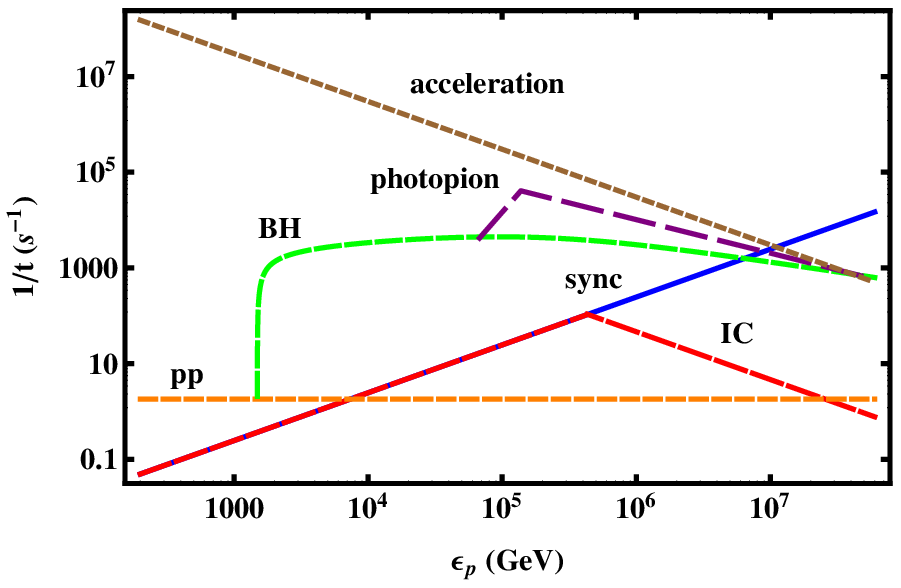}
\caption{Inverse of proton cooling and acceleration timescales (brown short dashed)
versus proton energy, in the jet head rest frame. Cooling mechanisms are synchrotron (blue solid), inverse Compton (red long dashed), proton proton collision (orange dashed), Bethe-Heitler (green dashed) and photopion process (purple dashed). Relevant
parameters: $L_j=10^{42}{\rm erg\,s}^{-1}$, $\theta_0=0.02$,
$r_{\rm He}=4\times10^{11}{\rm cm}$,
$\epsilon_e=\epsilon_B=0.1$. \label{Figure 1}}
\end{figure}

We now can plot the inverse of all these timescales as functions of
proton energy in Figure~\ref{Figure 1}. The maximum proton energy
can be obtained by $t_{\rm sync}=t_{\rm acc}$, so
$\epsilon_{p,\max}\simeq 1.11\times 10^7{\rm GeV}$. Hence we expect the maximum neutrino energy produced by this LL-GRB as $\epsilon_{\nu,\max}\simeq \frac{1}{10}\epsilon_{p,\max}\simeq 1.11{\rm PeV}$.
\par
We can define two threshold proton energies $\epsilon_{p,\rm
th}^{({\rm BH})}$ and  $\epsilon_{p,\rm th}^{(p\gamma)}$,
corresponding to $t_{pp}=t_{\rm BH}$ and $t_{\rm BH}=t_{p\gamma}$
respectively \citep{hor08}. We know that the proton-proton collision and
photopion process will produce muon neutrinos while the Bethe-Heitler
interaction does not. Hence, if the proton energy $\epsilon_p$ falls
into the range $\epsilon_{p,\rm th}^{({\rm
BH})}<\epsilon_p<\epsilon_{p,\rm th}^{(p\gamma)}$, the BH
interaction dominates and has a strong suppression on the final
neutrino spectrum. This suppression factor can be written as
$\zeta_{\rm BH}$,
\begin{equation*}
\zeta_{\rm BH}=\begin{cases}
\frac{t_{\rm BH}}{t_{pp}+t_{\rm BH}} & \text {if} \quad \epsilon_p<\epsilon_{p,\rm th}^{(p\gamma)},  \\
\frac{t_{\rm BH}}{t_{\rm BH}+t_{p\gamma}} & \text {if} \quad
\epsilon_p>\epsilon_{p,\rm th}^{(p\gamma)}.
\end{cases}
\eqno (17)
\end{equation*}

\par
Further on, the cooling of a meson also needs to be considered. It is
similar to  protons that the radiative and the hadronic cooling
times are
$$t_{\rm rad}=\frac{3m_\pi^4c^3}{4\sigma_Tm_e^2\epsilon(U_\gamma+U_B)},$$
$$t_{\rm had}=\epsilon/(c\sigma n_p \Delta{\epsilon}), \eqno (18)$$
where $\epsilon$ is in units of GeV, the numerical values are $t_{\pi,
\rm rad}=1.0\frac{1}{\epsilon_\pi}\,{\rm s}$, $t_{K,
\rm rad}=1.56\times10^2\frac{1}{\epsilon_K}\,{\rm s}$,  $t_{\pi,
\rm had}=t_{K,\rm had}=0.14\,{\rm s}$. Same as
\citet{hor08}, for our jet parameters, the meson goes from decay dominated to radiation cooling dominated. We can define the break energy for neutrinos,
$\epsilon_{\nu,\rm brk}$ satisfies $\gamma\tau\sim t_{\rm
rad}$, and thus the suppression factor due to meson cooling
is expressed as
\begin{equation*}
\zeta(\epsilon_\nu)=\begin{cases}
1 & \text {if} \quad \epsilon_\nu < \epsilon_{\nu,brk},  \\
\epsilon_{\nu,\rm brk}^{2}/\epsilon_\nu^2 & \text{if} \quad \epsilon_\nu \geq
\epsilon_{\nu,\rm brk}.
\end{cases}
\eqno (19)
\end{equation*}
\par
With the cooling suppression effect of both protons and mesons, we
obtain the final neutrino flux \citep{tot03, hor08}
$$F_\nu=\frac{\langle n \rangle B_\nu}{\kappa}\frac{L_{\rm iso}}{4\pi D_L^2
\ln(\epsilon_{p,\max}/\epsilon_{p,\min})}\frac{\zeta_{\rm
BH}(\epsilon_\nu)\zeta(\epsilon_\nu)}{\epsilon_\nu^2}, \eqno (20)$$
where $\langle n \rangle$ is the meson multiplicity (1 for pions and
0.1 for kaons),  $B_\nu$ is the branching ratio of meson decay into
neutrinos (1 for pions and 0.6 for kaons), and $\kappa^{-1}$ is the
fraction of the primary proton energy carried by neutrinos,
regardless of energy loss (1/8 for pions and 1/4 for kaons). The factor $\ln(\epsilon_{p,\max}/\epsilon_{p,\min})$ normalizes the proton spectrum to the jet power. In this paper, we just assume that the highly efficient acceleration occurs and take the acceleration
efficiency $\zeta_p\simeq 1$ so that $\epsilon_{acc}=\zeta_p(1-\epsilon_e-\epsilon_B)=0.8\sim 1$. This is consistent with the fiducial value of baryon loading parameter $\xi_{acc}\simeq \epsilon_{acc}/\epsilon_e\simeq 10$ according to \citet{mur07}.
\par
\begin{figure}[htbp]
  \centering
  \includegraphics[width=0.7\textwidth]{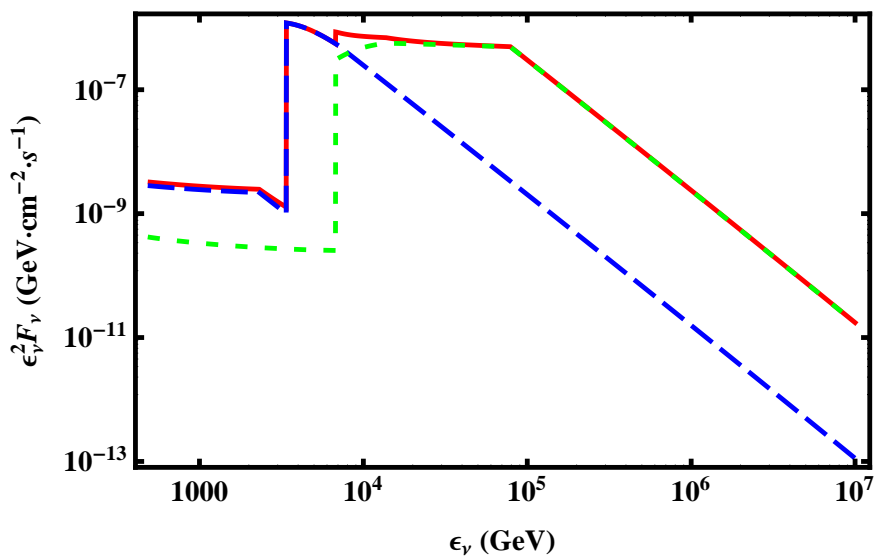}\\
  \caption{Neutrino flux multiplied by the square of neutrino energy versus neutrino energy for a LL-GRB.
  The blue dashed line represents the neutrino production through pion decay and the green
  dashed line is kaon decay. Meanwhile the red solid line is the total neutrino flux produced at jet head. Relevant parameters: $L_j=10^{42}{\rm erg\,s}^{-1}$, $\theta_0=0.02$,
  $r_{\rm He}=4\times10^{11}{\rm cm}$, $\epsilon_e=\epsilon_B=0.1$, at a distance $D_L=10\,{\rm Mpc}$. \label{Figure 2}}
\end{figure}

So we plot in Figure~\ref{Figure 2} the flux of
neutrinos for the LL-GRB jet mentioned above, with luminosity
$L_j=10^{42}{\rm erg\,s}^{-1}$, initial opening angle $\theta_0=0.02$. We assume that this LL-GRB is at a rather close
distance $D_L=10\,{\rm Mpc}$. We can see that at low energy end, $\epsilon_\nu^2F_\nu\sim {\rm const}$ suggests a power law neutrino spectrum. The neutrino
number from kaon decay is one or two orders of magnitude more than
that from pions at high energy end. It is mainly because kaons are heavier and
experience less energy loss\citep{hor08}.  A sharp jump is obvious in the spectrum
due to the transition of dominance from the BH interaction to photopion
process in proton cooling mechanisms thus prominently more neutrinos are produced, that is to say, it is caused by
$\zeta_{\rm BH}(\epsilon_\nu)$.
\par
We now can simply estimate the neutrino events for this LL-GRB in IceCube. We use the following fitting formula of the probability of detecting muon neutrinos \citep{mur06,ice11}.
$$P(E_\nu)=7\times10^5(\frac{E_\nu}{10^{4.5}\rm GeV})^\beta $$
where $\beta=1.35$ for $E_\nu < 10^{4.5}\rm GeV$, while $\beta=0.55$ for $E_\nu > 10^{4.5}\rm GeV$. The number of muon neutrinos from a burst are given by
$$ N(>E_{\nu,3})=A_{det}\int_{\rm TeV}^{\epsilon_{\nu,\max}}dE_\nu P(E_\nu)\frac{dN_\nu(E_\nu)}{dE_\nu dA}$$
Using a geometrical detector area of $ A_{det} = 1{\rm km}^2$, the expected neutrino number is $N\simeq 4.2\times10^{-3}$ for above LL-GRB with a neutrino emission duration of $T_{dur}\simeq r_{\rm He}/\beta_hc\simeq117\rm s$. Thus it is not easy to be detected now.

\par
\subsection{$1\ll \tilde{L} \lesssim \theta_0^{-4/3}$}

In this case, the jet is still collimated but the head velocity will
become subrelativistic. Similar to the above discussions, we deduce
some crucial terms below,
$$\beta_h \simeq 1,$$
$$\tilde{L}=4(2\pi)^{-2/5}L_j^{2/5}A^{-2/5}\theta_0^{-8/5}c^{-6/5},$$
$$\Gamma_h=\sqrt{\frac{1}{2}}\tilde{L}^{1/4}=(2\pi)^{-1/10}L_j^{1/10}A^{-1/10}\theta_0^{-2/5}c^{-3/10},$$
$$\bar{\Gamma}_h=\Gamma_j\sqrt{\frac{1}{2}}\tilde{L}^{-1/4}=2^{-9/10}\pi^{1/10}\Gamma_jL_j^{-1/10}A^{1/10}\theta_0^{2/5}c^{3/10},$$
$$\theta_j=2^{-4/5}\pi^{-3/10}L_j^{3/10}A^{-3/10}\theta_0^{4/5}c^{-9/10},$$
$$\rho_j=2^{8/5}\pi^{-2/5}L_j^{2/5}\Gamma_j^{-2}A^{3/5}r^{-2}\theta_0^{-8/5}c^{-6/5}. \eqno (21)$$
The internal energy of the shocked ambient medium by FS and the
shocked jet by RS are
$$e_f=(4\Gamma_h+3)(\Gamma_h-1)\rho_ac^2,\quad \quad e_r=(4\bar{\Gamma}_h+3)(\bar{\Gamma}_h-1)\rho_jc^2. \eqno (22)$$
\par
\begin{figure}[htbp]
  \centering
  \includegraphics[width=0.7\textwidth]{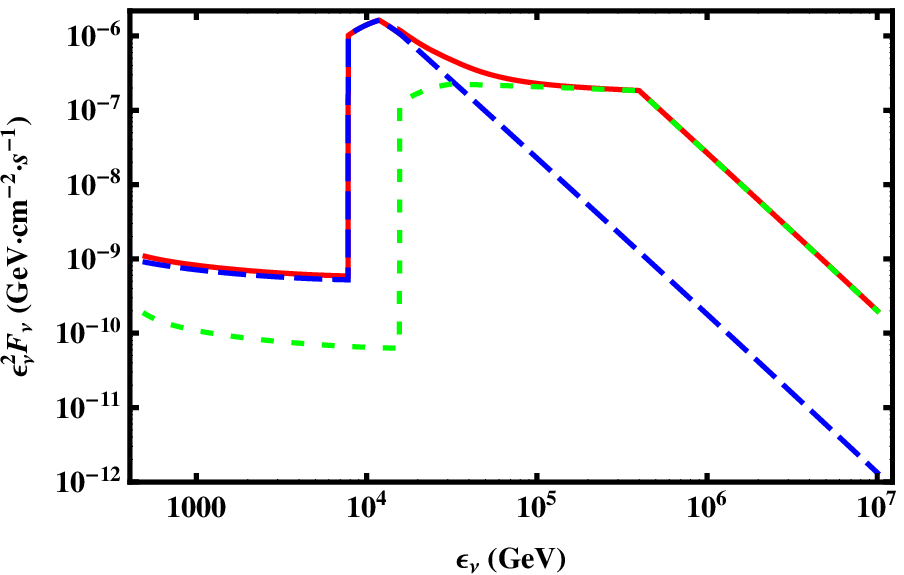}\\
  \caption{Neutrino flux multiplied by the square of neutrino
  energy versus neutrino energy, for the case of UL-GRB. The blue dashed line represents the neutrino production through pion decay and the green dashed line is kaon decay. The red solid line is the total neutrino spectrum produced at jet head. Relevant parameters: $L_j=2.5\times10^{44}{\rm erg\,s}^{-1}$, $\theta_0=0.01$,
  $r_{\rm He}=5\times10^{13}{\rm cm}$, $\epsilon_e=\epsilon_B=0.1$, at a distance
  $D_L=500\,{\rm Mpc}$. \label{Figure 3}}
\end{figure}

The protons will be accelerated simultaneously by FS and RS, but the forward shock's contribution is negligible for the reason that the FS would be radiation mediated, and the shock acceleration would be inefficient. In this case, we choose an ultra-long GRB with $L_{\rm iso}=10^{49}{\rm erg\,s}^{-1}$ and $\theta_0=0.01$, which suggests $L_j=2.5\times10^{44}{\rm erg\,s}^{-1}$ The extreme long duration $\sim 10^4$s suggests a progenitor like a blue supergiant (BSG) of radii up to $\sim 10^{13}{\rm cm}$. We assume this BSG with a helium core of mass $\sim 2M_\odot$ and radius $r_{\rm He}=5\times10^{13}\,$cm so that the envelope density can be expressed as $
\rho_a(r)=6.37\times 10^{18}r^{-2}\;{\rm g\,cm}^{-3}$. The assumed radius may be relatively larger than that of typical BSG, but we choose this value in order to realize efficient Fermi acceleration here. These parameters are possible according to \citet{woo12} and then we focus on the neutrino emission of this single UL-GRB. From equation (21) we get $\tilde{L}\simeq 14.06 \;,\;\rho_j=3.58\times 10^{-12}\;{\rm g\,cm}^{-3}\;,\;\Gamma_h\simeq1.37\;,\;\bar{\Gamma}_h\simeq36.5 $ thus this UL-GRB satisfies $1\ll \tilde{L} \lesssim \theta_0^{-4/3}$. The Thomson optical depth is $\tau_T=n\sigma_T l=\frac{\rho_j}{m_p}\sigma_T\frac{r_{\rm He}}{\Gamma_j} \simeq0.71(\frac{L_j}{2.5\times10^{44}{\rm erg\,s}^{-1}})^{\frac{2}{5}}(\frac{\theta_0}{0.01})^{\frac{7}{5}}(\frac{M_{\rm He}}{2M_\odot})^{\frac{3}{5}}(\frac{r_{\rm He}}{5\times10^{13}\rm cm})^{-\frac{8}{5}} \sim 0.1C^{-1}\bar{\Gamma}_h < 1$ so Fermi acceleration for RS is efficient. In line with the observation now (see \citet{gen13} ), we assume this UL-GRB at a relatively close distance $D_L=500\rm Mpc$. We calculate the neutrino flux for RS then we plot it in Figure~\ref{Figure 3}. The maximum proton energy $\epsilon_{p,\max}$ can be obtained by $t_{\rm sync}=t_{\rm acc}$, and the maximum neutrino energy produced by this UL-GRB is $\epsilon_{\nu,\max}\simeq\Gamma_h\frac{1}{10}\epsilon_{p,\max}\simeq 3.42{\rm PeV}$. Also, Greater total neutrino fluence is exhibited in this UL-GRB case. Same as before, the expected neutrino number in IceCube for this UL-GRB is $N\simeq8.3\times10^{-2}$, with a neutrino emission duration of $T_{dur}\simeq r_{\rm He}/\beta_hc\simeq2438 \rm s$.

\subsection{$\tilde{L} > \theta_0^{-4/3}$}
This case corresponds to the uncollimated regime. That is, the
collimation  effect is weak and to a good approximation the jet
remains conical. The head will move forward with an relativistic
velocity, $\beta_h=1$. In this case, we have
$$\tilde{L}=\frac{L_j}{\pi A\theta_0^2c^3},$$
$$\Gamma_h=2^{-1/2}\pi^{-1/4}L_j^{1/4}A^{-1/4}\theta_0^{-1/2}c^{-3/4},$$
$$\bar{\Gamma}_h=\Gamma_j\sqrt{\frac{1}{2}}\tilde{L}^{-1/4},$$
$$\theta_j=\theta_0,$$
$$\rho_j=\pi^{-1}L_j\Gamma_j^{-2}r^{-2}\theta_0^{-2}c^{-3}. \eqno (23)$$
\par
Unfortunately, this situation is not suitable for high energy
neutrino  production. The reason is that, to meet the requirement
$\tilde{L} > \theta_0^{-4/3}$, we need $L_j=10^{53}{\rm erg\,
s}^{-1}$ for $\theta_0=0.1$ in the helium core we previously
assumed. Leaving aside the existence of such a powerful jet, the Fermi acceleration is no longer efficient and
meson cooling in this jet is so severe that there will be hardly any
high energy neutrinos. Hence we do not need to carry on.

\section{Dependence on $\alpha$}
It is reasonable to argue that the final neutrino flux depends on the density
profile of progenitor envelope. Apparently, with the same assumed envelope mass and radius, different values of the power law index lead to different ambient envelope density, which directly influence the cocoon pressure and the collimation of the jet\citep{bro11}. Moreover, changing the power law index does affect the jet dynamics and result in different dependency on radius (we will see later). These two reasons cause the revision of collimation effect on the final neutrino spectrum being different. For the sake, we would like to study the influence of the different values of the power law index of the density profile. Here we display one situation (UL-GRB) for simplicity and clearness. We
still use the progenitor envelope properties of a helium core of mass $\sim
2M_\odot$ and radius $r_{\rm He}=5\times10^{13}{\rm cm}$, but with
$\alpha=2.5,\; 2.7$. Respectively, we can write them as
$\rho_a(r)=2.25\times10^{25}r^{-2.5}{\rm g\,cm}^{-3}$ and
$\rho_a(r)=7.42\times10^{27}r^{-2.7}{\rm g\,cm}^{-3}$. Surely, we
could repeat all those calculations above for $\alpha=2.5,\;2.7$.
\par
\begin{figure}[htbp]
  \centering
  \includegraphics[width=0.7\textwidth]{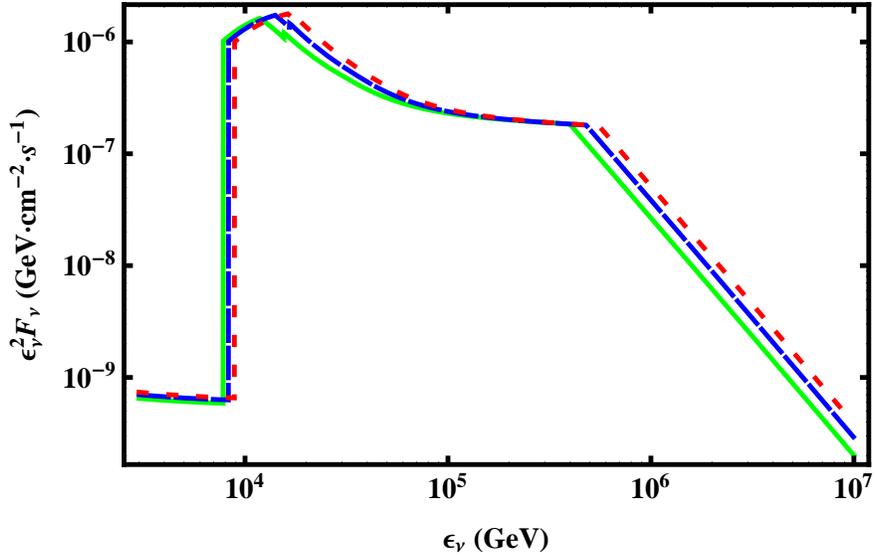}\\
  \caption{Total neutrino flux multiplied by the square of neutrino energy versus
  neutrino energy for a UL-GRB with different envelope density profile.
  Green solid lines, blue dashed lines and red dotted lines represent $\alpha=2$, $\alpha=2.5$, $\alpha=2.7$ respectively. Relevant parameters: $L_j=2.5\times10^{44}{\rm erg\,s}^{-1}$,
  $\theta_0=0.01$, $r_{\rm He}=5\times10^{13}{\rm cm}$, $\epsilon_e=\epsilon_B=0.1$,
  at a distance $D_L=500\,{\rm Mpc}$. \label{Figure 4}}
\end{figure}

For a successful jet in the condition $1\ll \tilde{L} \lesssim
\theta_0^{-4/3}$, we show the crucial, analytical quantities for
$\alpha=2.5$,
$$\tilde{L}\simeq2.85L_j^{2/5}A^{-2/5}r^{1/5}\theta_0^{-8/5}c^{-6/5},$$
$$\theta_j\simeq0.334L_j^{3/10}A^{-3/10}r^{3/20}\theta_0^{4/5}c^{-9/10},$$
$$\rho_j\simeq2.85L_j^{2/5}\Gamma_j^{-2}A^{3/5}r^{-23/10}\theta_0^{-8/5}c^{-6/5}, \eqno (24)$$
and for $\alpha=2.7$,
$$\tilde{L}\simeq3.66L_j^{2/5}A^{-2/5}r^{7/25}\theta_0^{-8/5}c^{-6/5},$$
$$\theta_j\simeq0.295L_j^{3/10}A^{-3/10}r^{21/100}\theta_0^{4/5}c^{-9/10},$$
$$\rho_j\simeq3.66L_j^{2/5}\Gamma_j^{-2}A^{3/5}r^{-121/50}\theta_0^{-8/5}c^{-6/5}. \eqno (25)$$
In all of three cases $\alpha=2,\;2.5,\;2.7$, we only consider the reverse shock's contribution to the final high energy neutrino flux and we plot it in Figure~\ref{Figure 4}. Same as before, we could get the maximum neutrino energy and it is $3.42{\rm PeV},\; 4.42{\rm PeV},\; 5.31{\rm PeV}$ respectively. Once we could correlate an observed high energy ($\sim \rm PeV$) neutrino precursor with a GRB in the future, we can constrain the envelope property of progenitor by maximum neutrino energy.
\par
We can see in this figure that $\alpha$ has a visible influence on the total neutrino flux, though not so prominent. Differences occur mostly at the high energy end. Also, the energy at which the neutrino spectrum peaks and the maximum neutrino energy in these three cases are a bit different. For a steeper envelope density profile, the density of the jet material and the
outer envelope at which neutrino production begins is generally
smaller. Hence, the steepest $\alpha=2.7$ case encounters the least cooling impact at high energy end thus leading to a highest maximum neutrino energy and a greatest neutrino flux. Anyway, we can make this dependency more striking, for example, by choosing $\alpha=2.9$. This could result in several times distinction compared to $\alpha=2$ case at high energy end. However, given the energy resolution of IceCube now, these differences are so small that we can hardly distinguish. Here we just provide an alternative way to probe the stellar structure and wish to do this if we could realize better energy resolution in the future.

\section{Comparison with Previous Works}
The main calculation of our paper is based on the jet propagation
dynamics developed by \citet{bro11}, and we further consider the
neutrino emission during the jet propagation process. This neutrino emission
serves as a precursor signal prior to GRB prompt emission. We differ from
\citet{mes01, enb09, mur13} at the point that we deal with the high energy
neutrino emission produced by shocks formed at the jet head, while
they focused on internal shocks. We have similar handle on the calculation
with \citet{hor08} but our result may be very different from theirs,
because they adopted a progenitor model from \citet{heg00} with a simplified
dynamics in which the jet opening angle remains constant and thus just
ignored the collimation effect, which should play an important role.
Collimation depends on the jet luminosity $L_j$, initial
opening angle $\theta_0$ and the progenitor density profile.
We calculate in detail the high energy neutrino flux in each
collimation regime, and choose the promising LL-GRBs and UL-GRBs as high energy neutrino sources, leading to a more likely result. And what is more,
we discuss the dependence of maximum neutrino energy and high energy neutrino flux on the
progenitor density profile.

\section{Discussions and Conclusions}
High energy neutrinos can be produced while the jet is still
propagating  in the envelope. These neutrinos appear as a precursor
signal, with energy ranges from GeV to PeV. Analytically, we
calculate this neutrino flux. To handle this, we first need to
determine whether the jet is in collimation regime. Collimation has
a crucial effect on jet propagation dynamics. We adopt the previous
propagation framework developed by \citet{bro11}.  We assume separated cases (LL-GRBs and UL-GRBs) in which Fermi first order acceleration works and they fall into different collimation regimes. With a power law spectrum of accelerated protons with $p=2$, then we calculate the neutrino flux in various situations.
\par
At low energy end, we always get $\epsilon_\nu^2F_\nu\sim {\rm const}$ and this suggests a power law neutrino spectrum. Neutrinos are mainly produced by proton-proton collision. As the neutrino energy goes higher, a sharp jump is obvious in the spectrum because of photopion process starting to dominate thus more neutrinos are expected. Moreover, the neutrino flux from kaon
decay is almost two orders of magnitude more than that from pions at high energy
end. It is mainly because kaons are heavier and experience less energy loss.
\par
In our expectation, the final neutrino flux will depend on the density
profile parameter $\alpha$. We take $\alpha=2,\;2.5,\;2.7$ to
verify this dependence. We get a good result in Figure~\ref{Figure
4}, which shows that the dependence is existing. Besides, the maximum neutrino energy for three cases is $3.42{\rm PeV},\; 4.42{\rm PeV},\; 5.31{\rm PeV}$ respectively. At a given radius,
the density of the jet material is lower for a steeper envelope
density profile. There is less cooling impact so that a higher maximum neutrino energy and a greater high energy neutrino flux is expected.
\par
In this paper, we only calculate the high energy neutrino flux
of one GRB for given parameters, but it is not easy to be detected
by the current instruments. We wish to obtain a diffuse GRB
neutrino background which can be correlated with current
observations in our future work.

\acknowledgments
We thank an anonymous referee for valuable suggestions that have allowed us
to improve this manuscript. This work is supported by the National Basic
Research Program of China (973 Program, grant 2014CB845800) and the
National Natural Science Foundation of China (grant 11033002).

\clearpage

\end{document}